\documentclass[sigplan,screen]{acmart}

\AtBeginDocument{%
  }

\usepackage{booktabs}
\usepackage{filecontents}

\usepackage{graphicx}
\usepackage[many]{tcolorbox}

\usepackage{mathtools,amssymb,latexsym,amsfonts,stmaryrd}
\usepackage{bbm}
\usepackage{pbox}
\usepackage{xfrac}
\usepackage{booktabs}
\usepackage{xcolor}
\usepackage{threeparttable}
\usepackage{amsthm}
\usepackage{float}
\usepackage{multirow}
\usepackage{tikz}
\usepackage{mathrsfs}
\usepackage{mathpartir}
\usepackage{subcaption}
\usepackage{algorithm}
\usepackage{algorithmicx}
\usepackage{url}
\usepackage{syntax}
\usepackage{framed}
\usepackage[noend]{algpseudocode}
\usepackage{flushend}
\usepackage{listings}
\usepackage{moresize}
\usepackage{wrapfig}
 
\usepackage{seqsplit}
\usepackage{alltt}
\usepackage{xspace}
\usepackage{enumitem}

\newcommand{\revise}[2]{{\color{red}{\ifx&#1&\else- #1\fi}} {\color{ForestGreen}{\ifx&#2&\else+ #2\fi}}}%
\renewcommand{\revise}[2]{#2}%

\usetikzlibrary{arrows,patterns, decorations.pathreplacing}

\newcommand{\ignore}[1]{}

\usepackage{pifont}

\usepackage{amsmath, listings, amsthm, amssymb, proof, xspace}
\usepackage{bbding}

\begin{document}

\title{Reasoning as a Resource: Optimizing Fast and Slow Thinking in Code Generation Models}

\author{Zongjie Li}
\email{zligo@cse.ust.hk}
\orcid{0000-0002-9897-4086}
\affiliation{%
  \institution{Hong Kong University of Science and Technology}
  \city{Hong Kong}
  \country{China}
}

\author{Shuai Wang}
\authornote{Corresponding author.}
\email{shuaiw@cse.ust.hk}
\orcid{0000-0002-0866-0308}
\affiliation{%
  \institution{Hong Kong University of Science and Technology}
  \city{Hong Kong}
  \country{China}
}

\begin{abstract}
This position paper proposes a fundamental shift in designing code generation models: treating reasoning depth as a controllable resource. Rather than being an incidental by-product of prompting, we argue that the trade-off between rapid, direct answers (``fast thinking'') and elaborate, chain-of-thought deliberation (``slow thinking'') must be explicitly managed. We contend that optimizing reasoning budgets across the entire model lifecycle—from synthetic data creation and benchmarking to real-world deployment—can unlock superior trade-offs among accuracy, latency, and cost. This paper outlines how adaptive control over reasoning can enrich supervision signals, motivate new multi-dimensional benchmarks, and inform cost-aware, security-conscious deployment policies. By viewing fast and slow thinking as complementary modes to be scheduled, we envision coding agents that think deep when necessary and act fast when possible.
\end{abstract}

\maketitle

\section{Introduction}
\label{sec:intro}

Large language models (LLMs) trained on code now exceed human-written baselines on benchmarks such as HumanEval and MBPP~\cite{HumanEval,MBPP}. Their practical utility, however, is constrained by a three-way tension among correctness, latency, and token cost. 
Recent hybrid models, exemplified by Qwen-3, expose explicit ``fast'' and ``slow'' modes that fuse direct answers with chain-of-thought traces, giving developers fine-grained control over reasoning at inference time~\cite{Qwen3Blog}. This development signals an important trend: reasoning depth is becoming a configurable parameter.

In this paper, we use the term \textbf{reasoning depth} to refer to the extent of deliberate, step-by-step problem-solving a model engages in before producing a solution. Chain-of-thought (CoT) traces—the externalized reasoning processes—serve as both the primary mechanism for implementing deeper reasoning and the measurable artifact that quantifies this depth in terms of token length and computational cost.

\begin{figure}[!htbp]
  \centering
  \includegraphics[width=\linewidth]{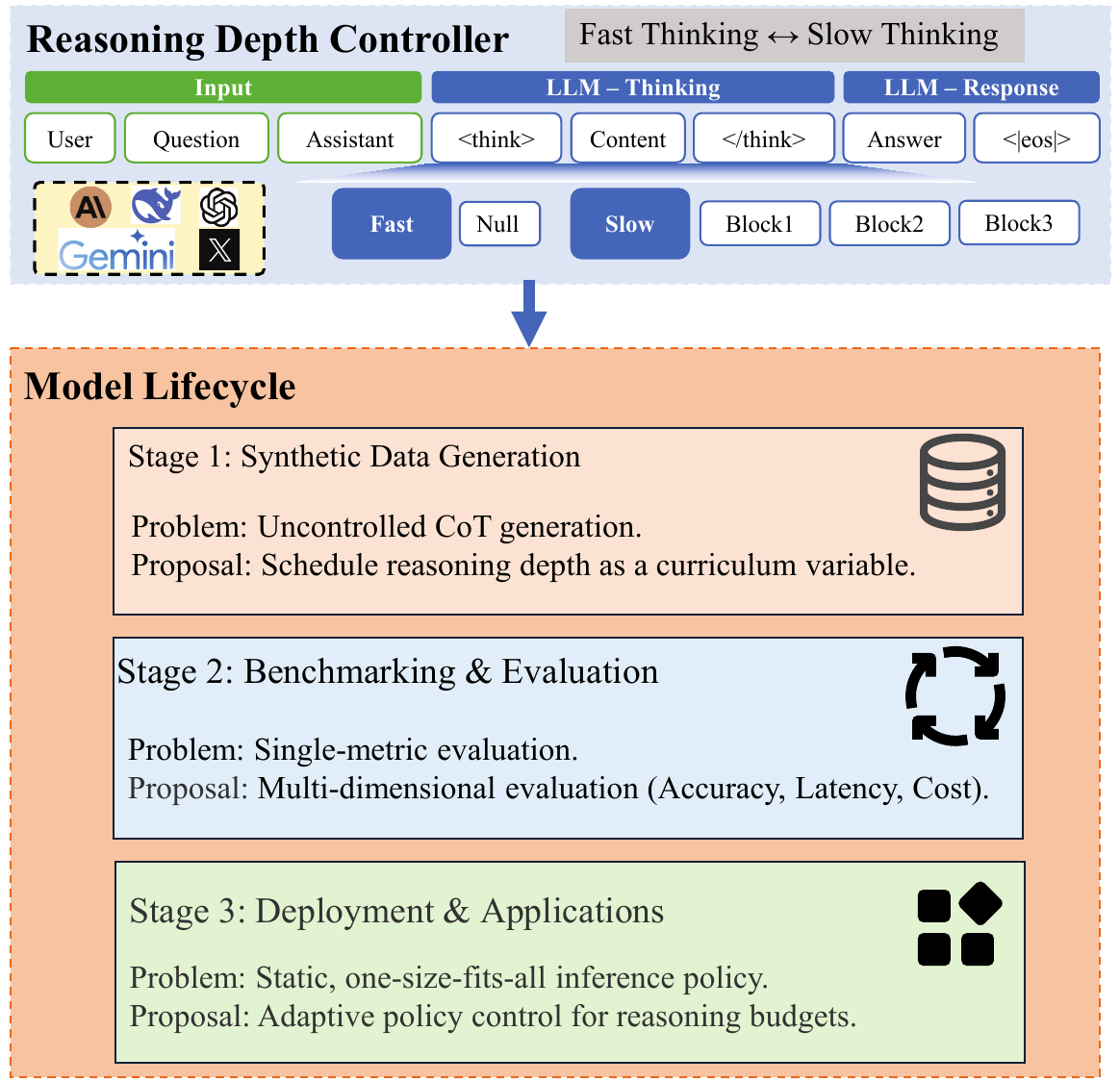} 
  \caption{An overarching framework for treating reasoning as a controllable resource. The top panel illustrates how the \textbf{Reasoning Depth Controller} modulates model behavior by choosing between ``Fast Thinking'' (direct answer generation) and ``Slow Thinking'' (generating an explicit chain of thought). The bottom panel depicts how this controller is applied to three critical stages of the \textbf{Model Lifecycle}: (1) strategically scheduling reasoning depth during \textbf{Synthetic Data Generation}; (2) enabling multi-dimensional evaluation---including accuracy, latency, and cost---during \textbf{Benchmarking \& Evaluation}; and (3) implementing adaptive reasoning budgets aligned with task and security policies during \textbf{Deployment \& Applications}.}
  \label{fig:overview}
\end{figure}

While adaptive schedulers like AdaCoT~\cite{AdaCoT25} and AdaptThink~\cite{AdaptThink25} can learn when deeper reasoning is warranted on a per-prompt basis, cutting inference budgets significantly without harming success rates, the broader research culture still optimizes reasoning depth in isolation. Synthetic-data pipelines often favor maximal CoT regardless of cost; benchmark suites measure only end-to-end accuracy; and production deployments seldom map service objectives to thinking budgets. Moreover, recent work on intellectual-property leakage in multi-agent systems reveals that unconstrained CoT can itself become an attack surface~\cite{wang2024benchmarking}.

We take the position that reasoning depth must be treated as a first-class design parameter throughout the entire lifecycle of code LLMs. Specifically, we contend that (i) synthetic-data generation should schedule CoT length to balance supervision richness against token economy, (ii) evaluation frameworks must report latency and token budgets alongside correctness, and (iii) deployment stacks require policy engines that allocate thinking budgets according to task difficulty and security constraints. In the following sections, we elaborate on each of these claims and outline concrete research opportunities.
Figure~\ref{fig:overview} provides a schematic of our proposed framework, illustrating how a central reasoning depth controller actively manages the trade-offs across the entire model lifecycle.

\section{Synthetic Code Data Generation}
\label{sec:synthetic}

We argue that synthetic data generation for code LLMs must explicitly optimize reasoning depth as a curriculum variable, balancing pedagogical richness against token economy and intellectual property protection.

The finite and uneven nature of publicly available source code has made synthetic data a necessity. Leading open models such as StarCoder are already trained on corpora where over half the tokens are machine-generated~\cite{StarCoder,TheStack}. This reliance on synthetic data is not a fringe practice but a prerequisite for scaling code LLMs beyond current limits.

To provide robust supervision, many pipelines embed \emph{reasoning traces} into the synthetic data, as code snippets alone are often insufficient. This approach has proven highly effective. Self-Instruct demonstrated that models can bootstrap their own instruction-response pairs for richer coverage~\cite{wang2022self}. For code, distillation methods that preserve intermediate problem-solving steps successfully transfer competence from large ``tutor'' models to smaller ``student'' models~\cite{DistillStepByStep23}. Further techniques enrich the supervision signal by guiding synthesis towards goals like improved API coverage~\cite{li2025api}, correcting specification misunderstandings~\cite{Tian2025MuFix}, incorporating structured information from compilers~\cite{li2022unleashing}, or even using adversarial setups to improve generation and search models synergistically~\cite{Wang2023GANCode}. Instruction-tuned models like Code Llama confirm that longer contexts mixing natural-language intent with step-wise rationale yield superior results~\cite{CodeLlama}.

However, this pedagogical richness comes at a significant cost. Reasoning depth directly inflates token budgets for data synthesis and subsequent model training. Reinforcement learning controllers like AdaCoT and AdaptThink address this by treating the decision to generate CoT as a learned action, invoking it only when beneficial~\cite{AdaCoT25,AdaptThink25}. Their success—recovering baseline accuracy with 50--70\% fewer tokens—demonstrates a Pareto improvement that could directly lower synthesis costs. Furthermore, long reasoning traces introduce security and IP risks. Each CoT may inadvertently reveal proprietary algorithms, test cases, or implicit specifications. Audits of multi-agent systems confirm that such traces can leak sensitive information even when final outputs are filtered~\cite{wang2024benchmarking}.

These observations imply that data-curation pipelines must move beyond a ``more is better'' approach to CoT. Scheduling reasoning depth as a curriculum variable, generating short traces for pattern-matching tasks but longer, detailed ones for complex algorithmic or security-critical examples, offers a principled path forward. This defines a concrete optimization target for the next generation of synthetic data systems.

\section{Inference-Time Testing \& Benchmarking}
\label{sec:benchmark}

We assert that code LLM evaluation frameworks must expand beyond pass@k metrics to report multi-dimensional trade-offs across accuracy, latency, and token usage, thereby capturing the full performance envelope.

The evaluation culture for code LLMs remains dominated by end-to-end correctness metrics like \textit{pass@k}, popularized by HumanEval and MBPP~\cite{HumanEval,MBPP}. While convenient, these scalar measures are insufficient because they mask the \emph{how} and the \emph{cost} of producing a solution. Two models can achieve identical scores even if one provides a concise, efficient answer while the other requires a lengthy, expensive chain of thought. This is a critical omission, as recent work on inference-time scaling laws shows that accuracy rises predictably with token count, implying a continuous trade-off between quality, latency, and budget~\cite{InferenceScaling24}. Taxonomies of code generation errors further reveal that failures often have complex root causes that simple pass/fail tests cannot capture~\cite{Wang2025UnderstandingErrors}. Yet, no mainstream code benchmark systematically reports on the reasoning process itself.

To expose these hidden performance dimensions, we propose that evaluation frameworks must log and report on the CoT traces. This enables a more nuanced analysis, as categorized in our diagnostic framework in Table~\ref{tab:matrix}. This matrix distinguishes between valid reasoning and correct solutions, providing deeper insights.
The ideal case is when correct reasoning leads to a correct solution. However, other quadrants are highly informative: correct reasoning that produces an incorrect solution may signal execution errors, while incorrect reasoning that coincidentally yields a correct solution points to potential overfitting or memorization—a critical concern for model robustness.

\begin{table}[ht]
  \centering
  \resizebox{0.99\linewidth}{!}{
  \begin{tabular}{p{3cm}|p{4cm}|p{4cm}}
  \toprule
  \multirow{2}{*}{\textbf{Reasoning Quality}} & \multicolumn{2}{c}{\textbf{Solution Correctness}} \\
  \cmidrule{2-3}
  & \textbf{Correct Solution} & \textbf{Incorrect Solution} \\
  \midrule
  \textbf{Correct CoT} & Sound understanding and implementation (ideal case) & Under-generalization or execution error \\
  \hline
  \textbf{Incorrect CoT} & Overfitting, memorization, or coincidental correctness & Comprehensive failure \\
  \bottomrule
  \end{tabular}
  }
  \caption{Diagnostic matrix for analyzing reasoning quality (via CoT traces) alongside solution correctness.}
  \label{tab:matrix}
\end{table}

This diagnostic framework allows researchers to differentiate models that genuinely solve problems from those that succeed on benchmarks through less reliable means. 
It also accounts for known issues in evaluation, such as position bias in LLM-based evaluators~\cite{li2024split} or subtle inconsistency failures in code completion~\cite{li2022cctest}. 
Techniques like oracle-guided program selection, which intelligently rank multiple candidates, further underscore the need for evaluation beyond a single-pass answer~\cite{Fan2024OracleGuided}.

Recent research reinforces the need for such reasoning-aware diagnostics. For instance, \emph{structured} prompting (e.g., SCoT), which encodes control flow into CoT, boosts accuracy with modest token overhead, proving that reasoning \emph{content}, not just length, matters~\cite{li2025structured}. Similarly, \emph{verify-and-edit} frameworks can correct reasoning chains post-hoc, but this is only possible if the chains are retained and exposed~\cite{zhao2023verify}. Studies on self-invoking generation also highlight the latency-accuracy trade-off inherent in longer traces~\cite{yu2024humaneval}.

Finally, as adaptive controllers like AdaCoT show, it is possible to match baseline pass@k with up to 70\% fewer tokens~\cite{AdaCoT25}. Without latency-aware benchmarks, these crucial Pareto improvements remain invisible. Just as traditional SAST tools are rigorously evaluated for their blind spots~\cite{li2024evaluating}, we must adopt a multi-dimensional perspective for code LLMs. We advocate for a new generation of evaluation suites that log timing, token counts, and reasoning traces by default, enabling us to quantify not only if an answer is right, but \emph{how efficiently and robustly it was produced}.

\section{Deployment and Downstream Applications}
\label{sec:deploy}

We maintain that production deployments of code generation models require adaptive reasoning-depth controllers that align thinking budgets with task complexity, service-level objectives, and security constraints.

Enterprise adoption has moved code LLMs into latency-critical and high-stakes environments, including IDE extensions, CI pipelines, and multi-agent systems. 
The applications are diverse and demanding, ranging from core software engineering tasks like automated program repair~\cite{wang2024navrepair, Eladawy24icse}, decompilation~\cite{wong2025decllm, wong2023refining}, and mutation testing~\cite{Deb2024SyntaxMutant}, to security-critical goals like repository-level auditing~\cite{Guo2025RepoAudit}, fuzzing~\cite{zhang2025low}, and even offensive security exercises~\cite{ji2025measuring}. In these settings, every token of reasoning incurs tangible GPU and monetary costs, making adaptive control of CoT a direct business concern~\cite{InferenceScaling24}. Controllers like \textsc{AdaCoT} and \textsc{AdaptThink} provide the mechanism for this, learning per-prompt policies that invoke CoT selectively to save tokens while preserving accuracy~\cite{AdaCoT25, AdaptThink25}.

The heterogeneity of tasks demands dynamic reasoning budgets. Generating a complex algorithm or a security exploit warrants deep deliberation, whereas autocompleting boilerplate code requires a near-instant response~\cite{MBPP}. Adaptive schedulers thus function as policy engines, mapping service-level objectives (e.g., 95th-percentile latency, max cost-per-request) to an appropriate CoT budget~\cite{li2025structured}.

Security considerations add a critical second dimension. The \textsc{MasLeak} framework shows that unconstrained CoT can leak proprietary logic in multi-agent systems~\cite{wang2024benchmarking}. This threat is compounded by risks such as training data extraction attacks~\cite{li2025differentiation} and model-stealing via imitation~\cite{li2023feasibility}. A robust deployment strategy must therefore treat CoT as a potential liability. Defensive measures should include not only prompt filtering but also CoT-aware sanitization, selective disclosure policies, and techniques like code watermarking to trace IP theft~\cite{li2023protecting, CodeLlama}. This must be part of a holistic security posture that also defends against adversarial inputs~\cite{wang2024selfdefend, wang2025stshield} and mitigates ethical inconsistencies or biases~\cite{ma2023oops, wang2024exploring}.

In sum, production-grade code LLMs demand a joint optimization of accuracy, latency, cost, and confidentiality. This cannot be achieved without elevating reasoning depth from an implicit behavior to a controllable resource, managed actively throughout the software lifecycle.

\section{Conclusion}
\label{sec:concl}

This position paper has argued that reasoning depth in code LLMs is not an incidental artifact but a \emph{first-class design parameter} that must be managed across data generation, evaluation, and deployment. We have contended that by actively scheduling chain-of-thought during data synthesis, reporting multi-dimensional trade-offs in benchmarks, and enforcing policy-driven reasoning controls in production, the field can build models that are both more capable and more efficient. The goal is to enable coding agents that \emph{think deep when necessary and act fast when possible}. We invite the programming languages and machine learning communities to embrace reasoning-budget awareness in future tools, frameworks, and optimization techniques, thereby unlocking a new Pareto frontier for practical and reliable code generation.

\bibliographystyle{ACM-Reference-Format}
\bibliography{./bib/code, ./bib/prosyn,./bib/sft,./bib/cot,./bib/imitation,./bib/ref,./bib/zjNewFull}

\end{document}